# Mitigating parasitic contributions in measured piezoresponse for accurate determination of piezoelectric coefficients in Sc-alloyed-AlN thin films using piezo-response force microscopy

Ch Kishan Singh,[1,a] K. Rajalakshmi,[a] N. Balamurugan,[b] Rakesh kumar,[c] Mukul Gupta,[d] R. Ramaseshan[a] and Kiran Baraik[c]

[a] *Surface and Thin-films Studies Section, Surface and Sensors Studies Division, Material Science Group, Indira Gandhi Centre for Atomic Research, A CI of HBNI, Kalpakkam – 603102, India*

[b] *Department of mechanical engineering, National Institute of Technology, Calicut - 673 601, India*

[c] *Accelerator Physics and Synchrotrons Utilization Division, Raja Ramana Centre for Advanced Technology, Indore, 452017, India*

[d] *UGC-DAE Consortium for Scientific Research, Khandwa Road, Indore - 452017, India*

## Abstract

We present a methodology to mitigate the effect of the parasitic electrostatic contribution usually present in piezoresponse force microscopy (PFM) measurement for quantitative characterization of polycrystalline piezoelectric thin films using a case study on a set of $Al_{1-x}Sc_xN$ thin films. It involves minimizing the voltage sensitivity of the measured piezoresponse by optimizing the optical lever sensitivity using the laser positioning of the beam-bounce system. Additionally, applying a *dc*-voltage offset (determined through Kelvin probe force microscopy) during PFM scans and positioning the probe over the interior/edge portion of the specimen are explored to minimize the local and non-local electrostatic tip-sample interaction. The results shows that the effective piezoelectric coefficient ($d_{33\text{-}eff}$) of our *c*-axis oriented wurtzite (*wz*)-$Al_{1.0}Sc_{0.0}N$ thin film is ~ 4.9 pm/V. The highest enhancement in the $d_{33\text{-}eff}$ value occurred in the *wz*-$Al_{0.58}Sc_{0.42}N$ thin film. Above $x > 0.42$, the $d_{33\text{-}eff}$ reduces due to phase-mixing of the *wz*-$Al_{1-x}Sc_xN$ phase with *cubic*-$Sc_3AlN$ phase till the piezoelectricity finally disappear at $x \approx 0.51$.

**Keywords:** Piezoresponse force microscopy, minimizing electrostatic artifacts in PFM, $Al_{1-x}Sc_xN$ thin films, piezoelectric coefficients

[1] Email: kisn@igcar.gov.in; kisnsingh@gmail.com

## 1. Introduction

The rapid miniaturization of almost all electronic devices in the past few decades have put a spotlight on the importance of researches involving low-dimensional systems like thin film, nanostructure, etc. In this regard, the field of piezoelectric and ferroelectric materials is no different as it has garnered a lot of interests due to their potential in device applications. The mandate highlighted a growing need for developing methods for accurate characterization of the piezoelectric properties of these low dimensional systems, particularly since the bulk techniques those are available do not work efficiently at the nanoscale as has been rightly pointed out by Kwon *et. al* [1]. The advancement in the field of atomic force microscope (AFM) has catered to this need with the development of voltage-modulation mode known as the piezoresponse force microscopy (PFM), which has now become the de-facto technique for electromechanical characterization of piezoelectric and ferroelectric thin films [2]. PFM utilizes the inverse piezoelectric effect (IPE) wherein an alternating current (*ac*) electrical field $V_{tip} = V_{ac}\mathrm{Sin}(\omega t)$ is transmitted to the specimen via a nano-sized conducting tip to excite it, and the electromechanical response from the specimen is measured with the help of a cantilever deflection detection system and a lock-in amplifier[3,4]. The cantilever deflection is detected using the optical lever of the beam bounce (OBB) method in most commercial AFMs. It works by detecting the reflection of a focused laser beam from the top of a cantilever's free end with a position-sensitive photodetector (PSPD). Here, it must be noted that the cantilever bending slope is measured rather than the deflection of the cantilever itself in AFMs using OBB method due to the sensitivity of the PSPD [5]. The cantilever bending slope is calibrated by scaling the output of PSPD against precise movements of the cantilever through the inverse optical lever sensitivity (*InvOLS*) and is very accurate when measured on a flat and hard surface [6].

The electromechanical response in a PFM measurement is expressed as the first harmonics of the tip oscillation $A_\omega \sin(\omega t+\varphi)$. The phase $\varphi$ provide information about the direction of polarization and the out-of-plane component of the amplitude $A_\omega$ is used to determine the effective piezoelectric coefficient $d_{33\text{-}eff}$ through the relation [7]:

$$A_\omega = d_{33-eff}\, V_{ac} \sin(\omega t) \quad (1)$$

However, the presence of several parasitic contributions in the measured $A_\omega$ complicate the accurate determination of $d_{33\text{-}eff}$. These non-piezoelectric contributions may arise from electrostatic interactions (local between the tip and sample; non-local between the Si-chip and sample), electrochemical strain, and other smaller effects like electrostriction, among others [7,8].

So, the measured amplitude function in Eqn. 1 can be generally written as $A_\omega = A_{piezo} + A_{elec} + A_{e\text{-}strain} + A_{others}$. Noting that the last contribution is quite small and insignificant (compared to the other three) that can be effectively accommodated or subsumed into the statistics, one sees that major non-piezoelectrical artifacts in $A_\omega$ can only stem from either the electrostatic or the electrochemical strain effects. Further, while the electrostatic contribution which is also larger in magnitude is always present in all the samples, the electrochemical strain contribution on the other hand arises only in the case of an ionically active sample [9,10]. Hence, in the present paper, we will focus on mitigating the effect of the ubiquitous contribution in the measured piezoresponse arising from the local and non-local electrostatic effects. In order to do so, we will briefly discuss the origin of this interaction and the prevailing methods that are available in the literatures to mitigate the same. The interaction between the tip and sample (usually dielectrics) in the presence of applied *ac* and *dc* ($V_{dc}$) voltages leads to a capacitive coupling whose Coulombic force can be expressed by the capacitance ($C$) gradient along the vertical direction ($z$) as

$$F_{elec} = k\,\delta_{elec} = \frac{1}{2}\frac{dC}{dz}\left(V_{dc} + V_{eff-pd} + V_{ac}\sin(\omega t)\right)^2 \qquad (2)$$

here, $k$ is the coupling spring constant, $\delta_{elec}$ is the effective displacement of the cantilever due to $F_{elec}$ and we defined $V_{eff\text{-}pd}$ as the effective electrostatic potential difference between the tip and sample arising from the cumulative effect of capacitive interaction, contact-potential and any possible charge injections. Then, upon adding the first harmonic component of the cantilever displacement due to $F_{elec}$ (see Eqn. 2), the amplitude response of Eqn. 1 modifies to

$$A_\omega = d_{33-eff}\,V_{ac}\sin(\omega t) + k^{-1}\frac{dC}{dz}\left(V_{dc} + V_{eff-pd}\right)V_{ac}\sin(\omega t) \qquad (3)$$

It is immediately evident from Eqn. 3 that using a cantilever with very high spring constant ($k \geq 40$ Nm) can significantly minimize the $\delta_{elec}$. However, the universal use of such stiff cantilever is impractical because of the detrimental effects the usage of high force can produce particularly in soft and ultra-thin film materials [11]. Moreover, it does not completely eliminate the second term of Eqn. 3. Hong *et al.,* minimizes the non-local component of the electrostatic interaction by scanning near the sample's edge so that there is minimal overhanging of the Si-chip over the sample surface [12]. Their approach however fails to address the local component of the electrostatic interaction between the tip and the sample. Another common approach for minimizing the electrostatic effect involves the application of an appropriate *dc* offset through $V_{dc}$ in Eqn. 3 [7]. This approach involves the use of either contact ($dc$-sweeping) or non-contact Kelvin probe force microscopy (KPFM) to estimate the $V_{cpd}$ [11,13]. Recently, Signore *et. al.,* has applied the $dc$-sweeping method to estimate the $V_{cpd}$ and use it as an offset during the piezo-characterization of AlN thin films [14]. It may be noted that the $V_{cpd}$ values reported in their study varies over 100% when the $V_{ac}$ was swept. This may be due a shortcoming common to all the $dc$-offset based (KPFM) approaches. These approaches have been focusing only on the tip-sample interaction. The contribution arising in the measured $A_{elec}$ or $A_{\omega}$ from the laser positioning of the optical lever system may have been overlooked so far. In fact, Killgore *et al.,* has recently reported that it is possible to suppress the effect of the electrostatic forces on the cantilever deflection and achieve a better quantification of piezoelectric coefficients by simply optimizing the laser positioning of the OBB [15]. They have achieved that by positioning the slope sensitive laser of the OBB along the length of the cantilever at a location least affected by the electrostatic forces. However, this approach too has a disadvantage due to inherent experimental uncertainties. If we fail to precisely position the laser at the desired spot, the measured piezoresponse will continue to be influenced by the remanent electrostatic effects. Such experimental uncertainties may occur due to various factors like sizes of the laser spots, shape and size of different cantilevers and human errors, etc. Under such circumstances, considering the $V_{dc}$ offset based elimination given by Eqn. 3 will be useful, but is still unexplored at the moment. From all the above discussions, it is clear that significant efforts have been put by several independent researchers to address the issue of parasitic artifacts plaguing the accurate quantification of piezoelectric coefficients using different approaches. However, a unified approach that address this issue by capitalizing on the positives of all the above approaches is still missing in literature and we strongly feel that it needs to be explored. Such an approach can ensure repeatability and consistency while using PFM for quantitative characterization of piezoelectric thin films.

In this report, we present a methodology that aims to mitigate the parasitic contribution to the measured piezoresponse of a piezoelectric thin film using the case study on a set of $Al_{1-x}Sc_xN$ thin films. The reason for choosing $Al_{1-x}Sc_xN$ thin films in our study were two-fold. First, to eliminate or minimize any contribution from electrochemical or Vegard strain effect in the measured piezo-response. Secondly, Sc-alloyed AlN has shown the most promising results of enhancement in the effective piezoelectric coefficient $d_{33\text{-}eff}$, among all the transition-metal-alloyed AlN experimentally explored so far [16].

**2. Experimental:**

We have deposited the set of pristine-AlN ($Al_{1.0}Sc_{0.0}N$) and $Al_{1-x}Sc_xN$ thin films on a conducting B-doped $p^+$-Si (100) substrates using reactive co-sputtering of 5N pure targets of Al and Sc at room temperature in a commercial *dc* magnetron sputtering unit (AJA Int. Inc.). The base pressure of the unit was ~ $7\times10^{-8}$ Torr, and we passed the reactive ($N_2$) and the sputter (Ar) gases at 37.5 and 12.5 sscm, respectively, to maintain a working pressure of ~ $2.9\times10^{-3}$ Torr during all depositions. We varied the power applied to the Al and Sc-targets in the range from 133 to 200 W and 0 to 200 W, respectively, to vary the Sc-content ($x$) in the $Al_{1-x}Sc_xN$ alloy thin film. In addition, we suitably adjusted the overall run-time of each deposition (with different powers to the Al and Sc-target) to ensure that all the $Al_{1-x}Sc_xN$ thin films had similar thickness of ~ 200 nm. The thicknesses of the $Al_{1-x}Sc_xN$ thin films thus achieved were measured using a stylus profilometer to be 200 ± 20 nm. We used a Bruker Discover D8 diffractometer equipped with a rotating Cu-anode source (λ = 1.5416 Å) operated at 4.5 kW for the X-ray diffraction (XRD) and grazing-incidence-XRD (GIXRD) measurements of all the thin films. A 0.5° angle of incidence was used during the GIXRD measurements. The compositional characterization of the $Al_{1-x}Sc_xN$ alloy thin films were investigated by X-ray photoelectron spectroscopy (XPS) using a 715eV excitation energy at Angle Resolved Photoelectron Spectroscopy (ARPES) beamline (BL-10), Indus-2 synchrotron radiation facility located at RRCAT, Indore.

We have used an NTEGRA AFM (NT-MDT, Spectrum Instruments) for performing all the PFM and the KPFM measurements presented in this study. A cantilever with moderate stiffness ($k$ ~ 6 N/m) having a Pt-coated conducting tip was used for both the measurements. The KPFM measurement was performed in the semi-contact mode using the first resonance ~ 114 kHz in a two-pass scheme with the *dc* voltage ($V_{dc}$) bias applied through the tip. The lift height was kept constant at 100 nm for all the samples. The PFM measurements were performed in contact-mode with the *ac* voltage ($V_{ac}$, 1 to 5 V) applied to the tip at a modulation frequency of 20 kHz, which was far below the tip-sample contact resonance frequency (CRF) of ~ 500 kHz (given in supplementary information), to prevent any topographical crosstalk. In order to optimize the electrostatic voltage sensitivity of the optical lever system, the piezoresponse was measured by modulating the AlN thin film by a 5V *ac* along with the application of a $V_{dc}$ sweep from -5V to +5V through the tip. The above procedure was repeated for various focusing point of the laser beam (of the OBB) along the length of the cantilever.

## 3. Results and discussion:

### *3.1 Phase and composition:*

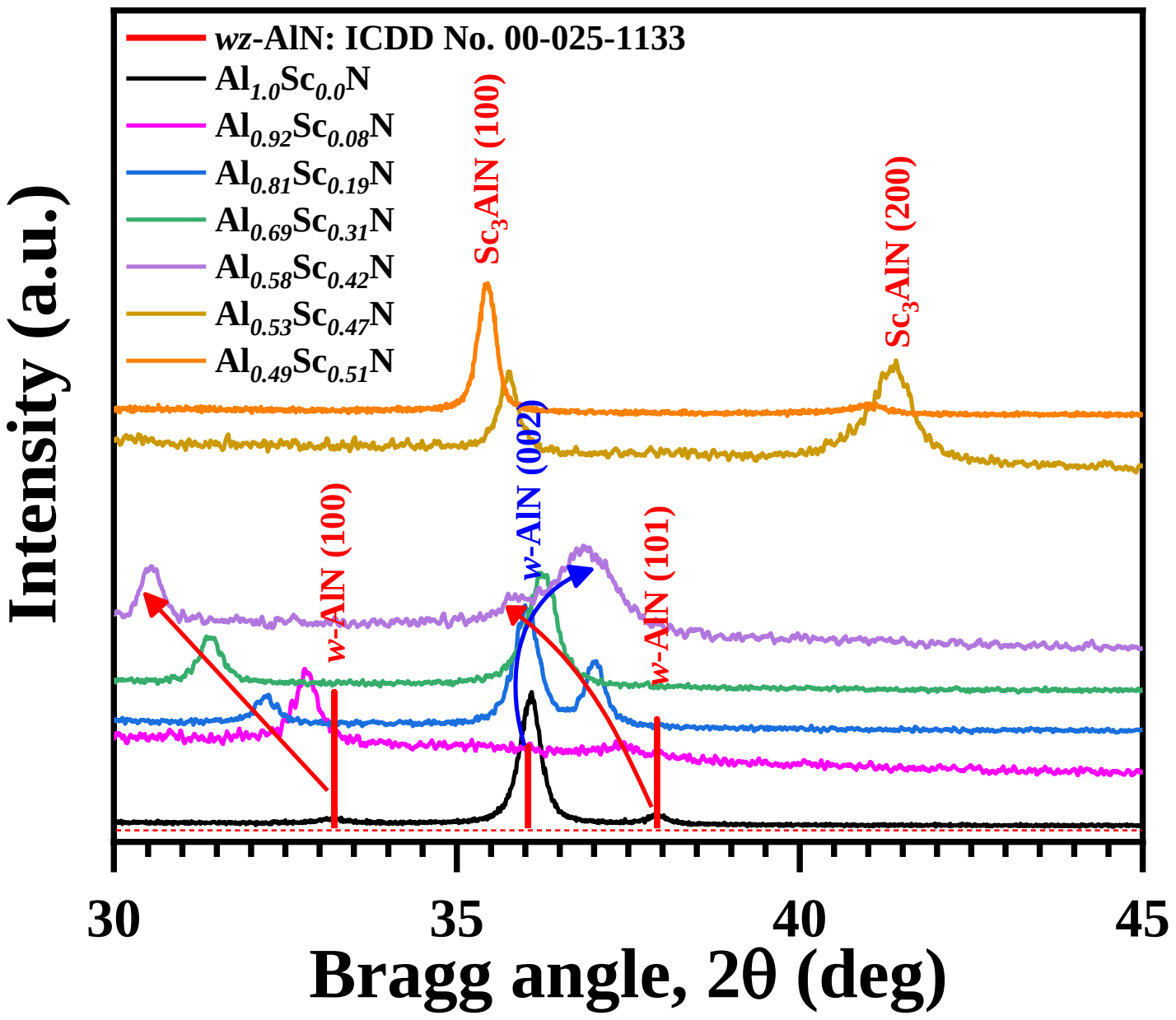

**Fig. 1:** GIXRD patterns of the various sputtered deposited $Al_{1-x}Sc_xN$ thin films.

The XRD patterns showing the phase information of the $Al_{1-x}Sc_xN$ thin films are shown in Fig. 1. We observe that the $Al_{1.0}Sc_{0.0}N$ (hereafter written as “AlN” for brevity) thin film crystallizes in the polycrystalline hexagonal-wurtzite (*wz*)-AlN phase (ICDD PDF No. 00-025-1133) and exhibit a strong orientation along the *c*-axis i.e., (002) reflection. Subsequent alloying of the AlN with Sc distorts the *wz*-AlN lattice, as indicated by the observed shifts in the (100), (002), (101) reflections. However, the $Al_{1-x}Sc_xN$ thin films, can be seen to retain the *wurtzite* phase till $x$ up to ≈ 0.42. Above $x \approx 0.42$, *wz*-$Al_{1-x}Sc_xN$ phase starts phase mixing with (and subsequently phase transform to) the cubic $Sc_3AlN$ phase (ICDD card no. 00-061-0104), in agreement in previous results [17]. The stoichiometry of our $Al_{1-x}Sc_xN$ thin films given in the legends of Fig. 1 and are based on composition estimated from photoelectron peaks in the XPS spectra (Fig. 2) of $Al_{1-x}Sc_xN$ thin films. In Fig. 2, the photoelectron peaks at binding energies (BE) ~ 75 eV, ~ 403 eV to ~ 408 eV and ~ 397 eV originated, due to photoemission from the Al 2p, Sc 2p doublet (Sc $2p_{3/2}$ and Sc $2p_{1/2}$) and N 1s core levels, respectively [18,19]. The Al 2p and Sc 2p doublet peaks are chemically shifted to higher BE along with a concomitant shift of the N 1s peak to lower BE indicating the co-ordination of Al─N and Sc─N bonds in the $Al_{1-x}Sc_xN$ thin film. The broad peak appearing at BE ~ 386 eV result from the Sc *LMM* Auger emission and the weak peak at ~ 414 eV is the satellite structure from the Sc 2p-doublet. Herein, we highlight that there is an overlap between the N 1s peak with the Sc $2p_{3/2}$ and the Sc *LMM* peaks, particularly at higher concentration of Sc. Hence, the stoichiometry of our $Al_{1-x}Sc_xN$ thin films given in Fig. 2 were estimated by normalizing the Al 2p and Sc 2p doublet peaks, after correcting for the relevant photo-electric cross-sections and inelastic mean free path, etc [20–22].

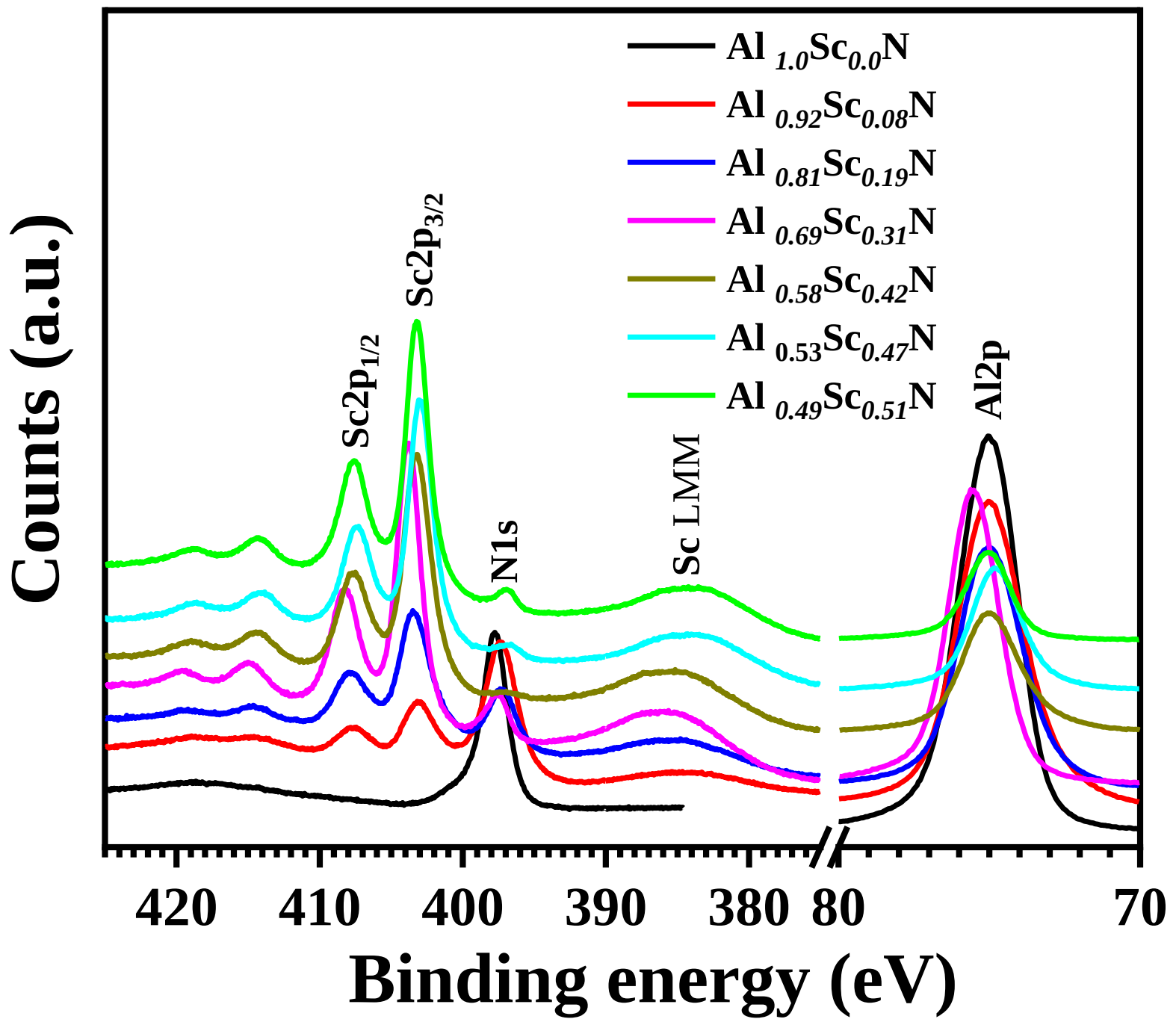

**Fig. 2:** XPS spectra showing the Al 2p, Sc 2p and N 1s peaks of various $Al_{1-x}Sc_xN$ thin films.

### *3.2 Piezoresponse force microscopy:*

The variation in the piezoresponse amplitude, measured in current (pA), of the pristine-AlN thin film modulated by a 5V *ac* due to *dc* voltage sweeps ($V_{dc}$ = ± 5V), when the laser spot used for optical beam-bounce (OBB) is sequentially focused at different indicated locations of the cantilever are shown in Fig. 3. We maintain a constant tip-sample force of interaction in all measurements using the force-distance (*F-d*) curve and the corresponding *InvOLS* during each sweep through the Hooke's law given as [23]:

$$F = k\,\delta = k\,I\,InvOLS \tag{4}$$

where $k$ is the spring constant of the cantilever, $\delta$ is the cantilever deflection and $I$ is the scaling function of the PSPD output. Fig. 3 shows that the position of the laser spot on the cantilever beam has a profound impact on the measured piezoresponse.

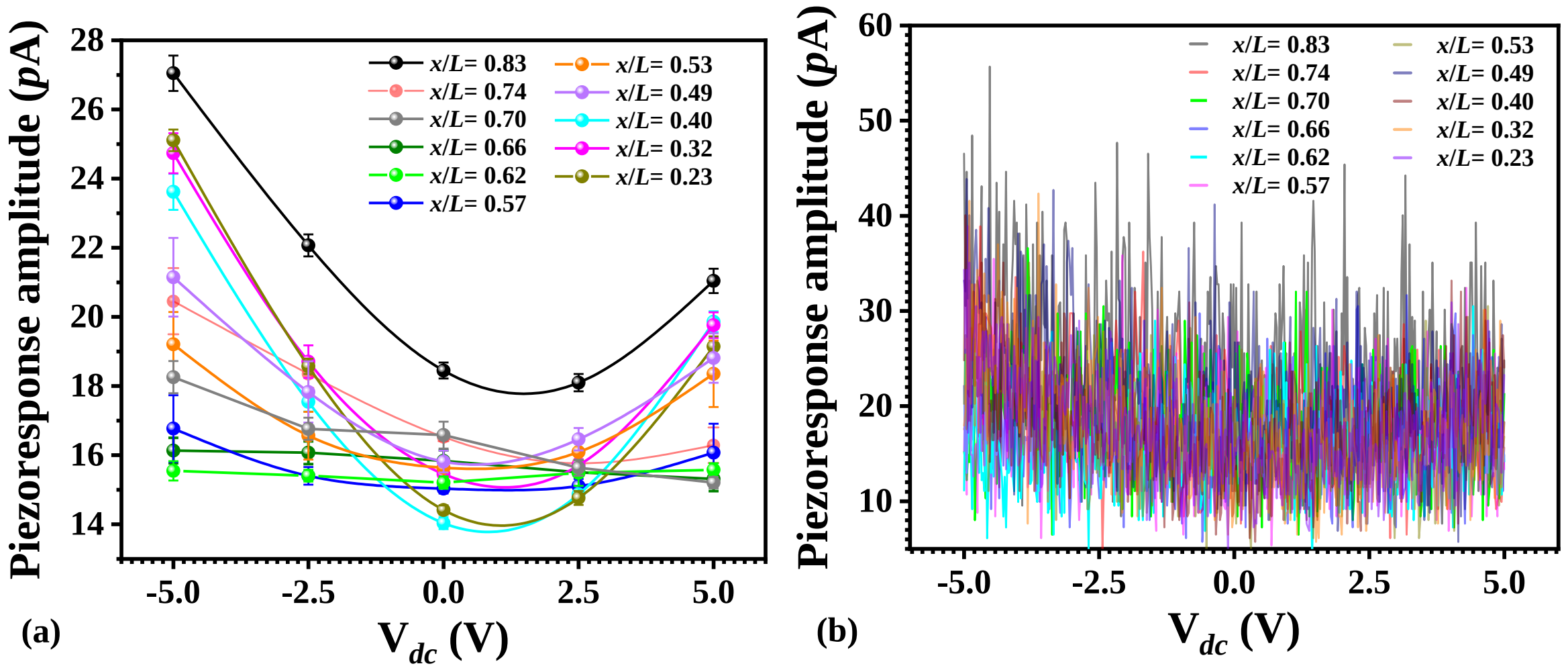

**Fig. 3:** Piezoresponse of the AlN thin film modulated by $V_{ac}$ (5V) to *dc* voltage ($V_{dc}$) sweep (± 5V) when the laser spot is focused at different positions along the length of the cantilever: Area scan (a) and corresponding Point spectroscopy (b).

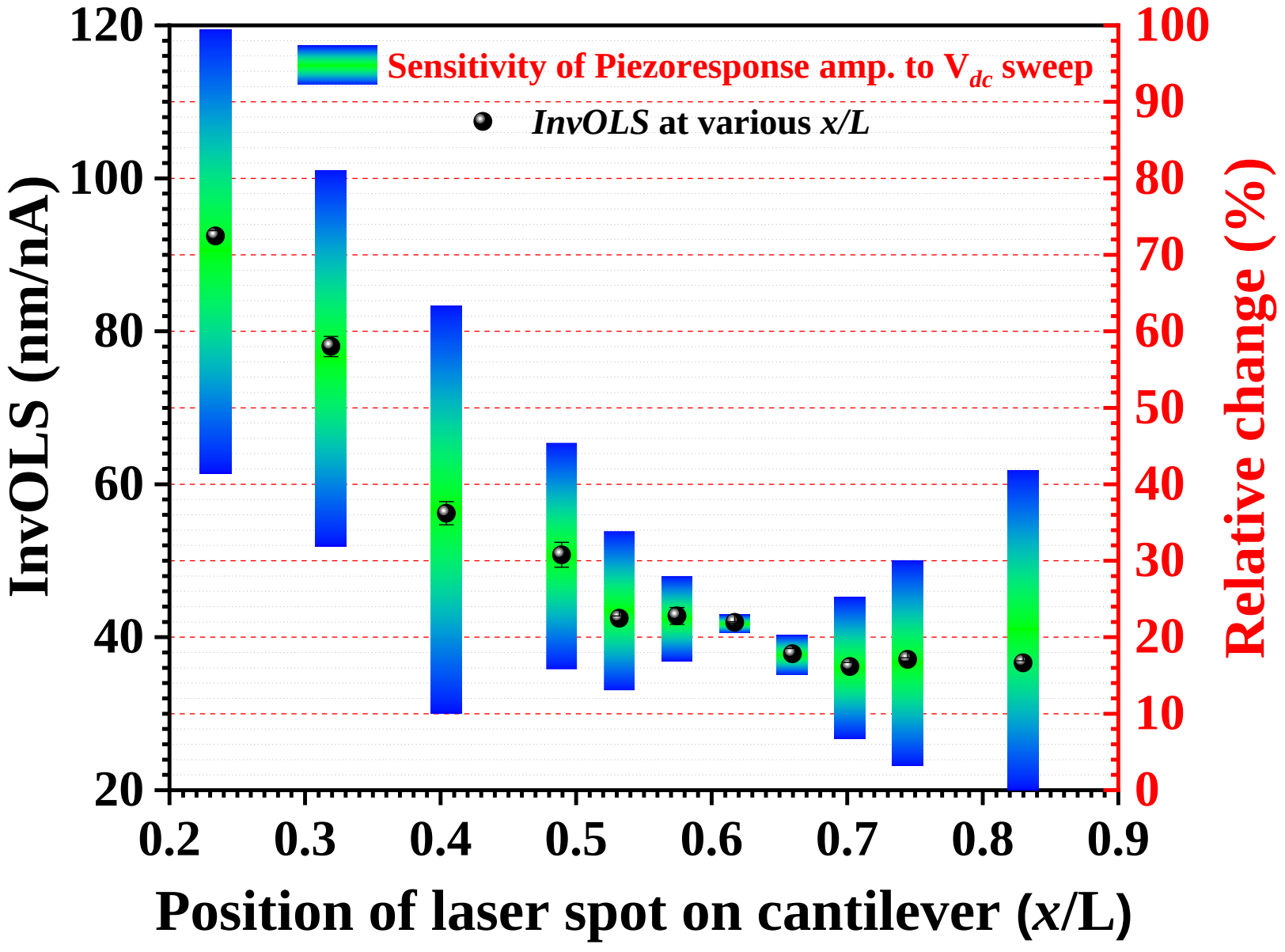

**Fig. 4:** The plot of the *InvOLS* and sensitivity of the piezo-response amplitude modulated by $V_{ac}$ (5V) to $V_{dc}$ sweeps *versus* the position of the laser spot along the length of the cantilever. *x* is measured from the base or fixed end of the cantilever.

Hence, in order to locate the position of the laser spot wherein the measured piezoresponse exhibit least sensitivity to the applied $V_{dc}$, we have plotted the relative change in the value of the measured piezoresponse (in *pm*) at various focusing point (*x,* measured from the base) of the laser spot along the length (*L*) of the cantilever in Fig. 4. This relative change (%) gives a measure of sensitivity of the measured piezoresponse to the overall effect of the applied $V_{dc}$ and any parasitic electrostatic effects that the tip-sample interaction may introduce to the system. We observe that the piezoresponse has a least sensitivity (~ 2.3 %) to $V_{dc}$ when the laser spot is focused at $x$/L ~ 0.617. This value is close to the normalized spot value of $x/L_{eff}$ ~ 0.6 reported by Naeem *et al.* at which the sensitivity of the cantilever deflection is optimum and independent of the cantilever's end load i.e., coupling spring constant [24]. A similar result was also reported by Killgore *et. al.,* where they modelled the cantilever as an Euler-Bernoulli beam [15,25]. The studies showed that while the cantilever bending due to piezoresponse was present all along the length of the cantilever as long as the tip-sample coupling is stiff, the bending due to electrostatic contribution had a null at $x$/L ~ 0.63 [15,25]. Hence, the *F-d* curve and the *InsOLS*, thus obtained at this optimized laser position of $x$/L ~ 0.617, can yield quantitatively accurate value of cantilever displacement due to IPE. Here, we stress that such optimization of the optical lever sensitivity is equivalent to stiffening the spring constant of the cantilever that helps to reduce the electrostatic contribution in $A_\omega$ through Eqn. 3. In PFM, the AFM cantilever is end loaded with the contact force applied through the tip. The $k$ of a cantilever with a rectangular cross section (width '$w$' and height '$h$' ) like the one we have used in our study has been experimentally observed to follow an inverse cube relation with its length ($L$) through the relation $k = 0.25Ewh^3/L^3$ [5,26]. So, when the laser of the OBB is moved away from the tip toward the base while maintaining a constant tip-sample force of interaction, the effective length of the cantilever decreases. Hence, the $k$ of an AFM cantilever normally specified at the free end (containing the tip) effectively becomes $k^*$ ~ 4.25 $k$ when the laser spot is placed at $x/L$ ~ 0.617. In our case, since we have used a cantilever with $k$ ~ 6 N/m, so our effective $k^*$ is ~ 25.5 N/m at $x/L$ ~ 0.617. This value in itself is close to the value of $k$ ( ~ 30) above which electrostatic contribution in $A_\omega$ has little effect [13].

After optimizing the $V_{dc}$ sensitivity of the optical lever system via laser positioning, we used semi-contact KPFM to estimate an effective *dc* offset to eliminate any remanent electrostatic effects due to the uncertainties discussed earlier. This offset should eliminate the local-electrostatic effect between the tip and the sample. In addition, we also explore performing PFM scans at the edge of the sample to remove any remanent non-local electrostatic effect between the body of the cantilever and the sample. After all these considerations, we have performed PFM measurements on our thin film samples employing four different configurations schematically illustrated in Fig. 5(a) to (d). In all the four set of experiments, we have focused the laser spot of the OBB at $x$/L ~ 0.617 and apply the $V_{ac}$ (1 to 5 V) through the conducting tip at 20 kHz. In Fig. 5(a), we performed the $V_{ac}$ sweep deep inside the sample such that the Si-chip supporting the cantilever overhang the thin film sample by a large margin.

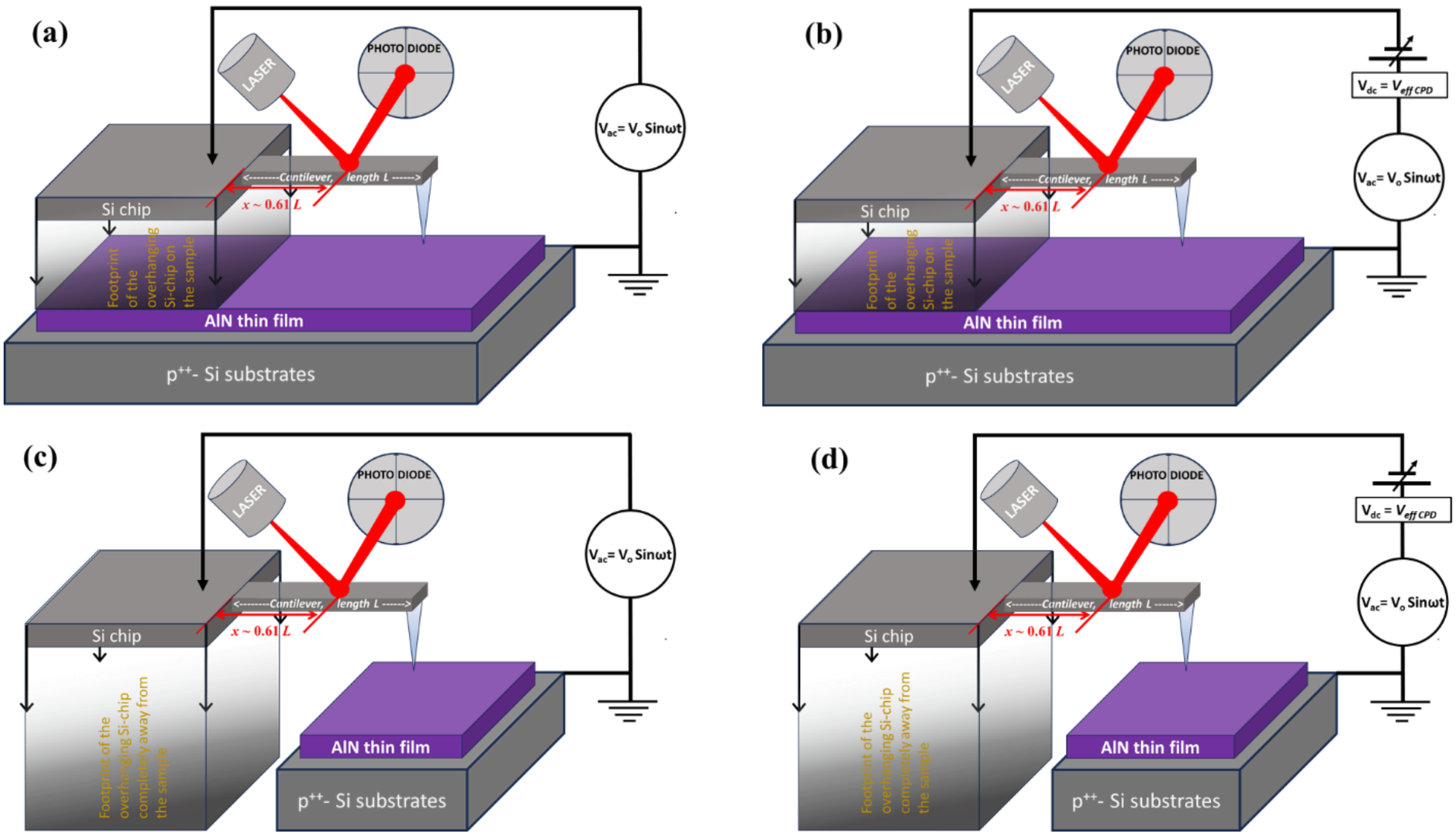

**Fig. 5:** Schematic illustration of the different configurations used for PFM measurements: (a) Configuration-1: the Si-chip and the cantilever overhangs the sample surface, (b) Configuration-2: similar to configuration-1 except the application of an additional $V_{dc\text{-}eff}$, (c) Configuration-3: the probe is moved to the edge of the sample such that only the unavoidable portion of the cantilever overhang the sample surface, and (d) Configuration-4: similar to configuration-3 except the application of an additional $V_{dc\text{-}eff}$. Please note that we have focused the laser spot of the OBB at x ~ 0.617 L and applied the $V_{ac}$ (1 to 5V) to the tip in all the configurations

Configuration-1 will reveal the efficacy of focusing the laser spot at $x$/L ~ 0.617 in suppressing the electrostatic artifacts that the overhanging Si-chip produces in presence of the applied field. Next, the configuration-2 shown in Fig. 5(b) is similar to configuration-1 except for the application of an additional *dc* bias to offset any effective contact potential that is not minimised in configuration-1 during the $V_{ac}$ sweep. This *dc* offset was determined by KPFM measurements on the sample by varying the $V_{ac\text{-}ex}$ (excitation of the 2$^{nd}$ pass) from 1 to 5V to exactly match the values of $V_{ac}$ subsequently used for sweeping in our PFM measurements. The value of the measured offset remained constant within the experimental error limit when the $V_{ac\text{-}ex}$ was varied from 1 to 5V, and we have given their values for different samples in the supplementary information. Then, in the configuration-3, we move the probe to the edge of the sample such that the Si-chip does not overhang the thin film specimen during the $V_{ac}$ sweep as shown in Fig. 5(c). Finally, configuration-4 is similar to configuration-3 except for the application of an additional *dc* bias determined through KPFM measurements similar to the case of configuration-2.

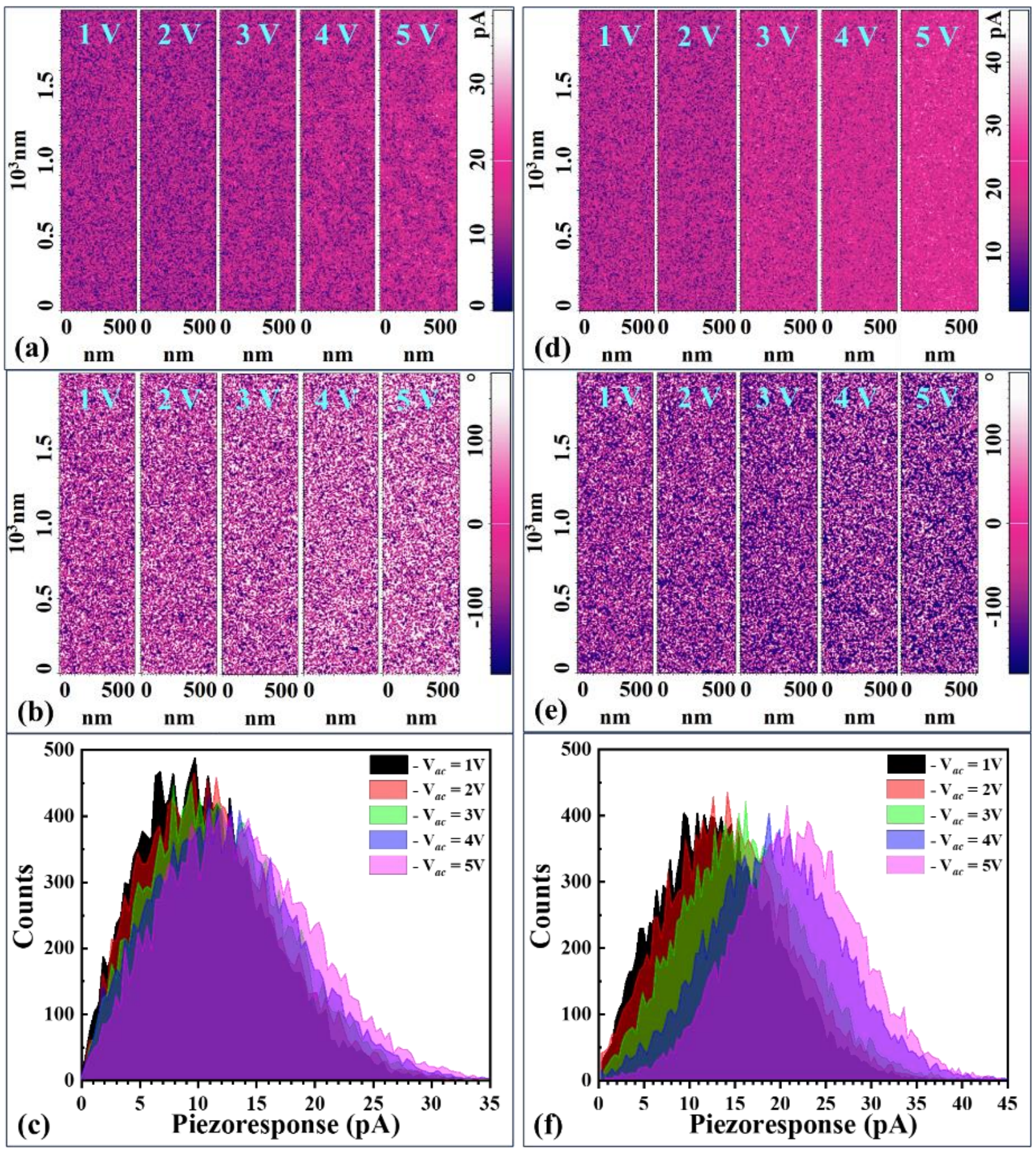

**Fig. 6:** Piezoresponse amplitude, phase image and histogram plots of: AlN thin film {(a) to (c)}, and $Al_{0.58}Sc_{0.42}N$ thin film {(e) to (f)}, obtained with a sequential $V_{ac}$ sweeping of the same scanning area using configuration-1.

Thereafter, in each configuration also, we have acquired the $V_{ac}$-sweeping data using two different modes of data acquisition for better statistics. To illustrate this, we have presented results from $V_{ac}$-sweeps of two representative specimens (AlN and $Al_{0.58}Sc_{0.42}N$ thin films) obtained in configuration-1 using the two different modes of data acquisition in Fig. 6 and Fig. 7. For the first mode of data acquisition, Fig. 6(a), (b) and (c), respectively, show the piezoresponse amplitude (in pA), the corresponding phase images and the histogram plots, of the AlN thin film wherein the $V_{ac}$ is sequentially sweep (from 1 to 5 V) over the same scanned area repeatedly. Similarly, Fig. 6(e), (f), and (g), respectively, show the same set of results from the sequential $V_{ac}$-sweeping over the same scanned area for the $Al_{0.58}Sc_{0.42}N$ thin film. For both the specimens, a systematic increase in the piezoresponse amplitude with an increase in the value of the applied $V_{ac}$ could be clearly observed from various amplitude images and shifts in their corresponding histograms shown in Fig. 6. The mean value of the histogram of the piezoresponse amplitude increases linearly from 10.7 pA at 1 V to 13.8 pA at 5 V for the AlN thin film. In the case of $Al_{0.58}Sc_{0.42}N$ thin film, the mean value of histogram of the piezoresponse amplitude increases linearly from 14.4 pA at 1 V to 27.8 pA at 5 V.

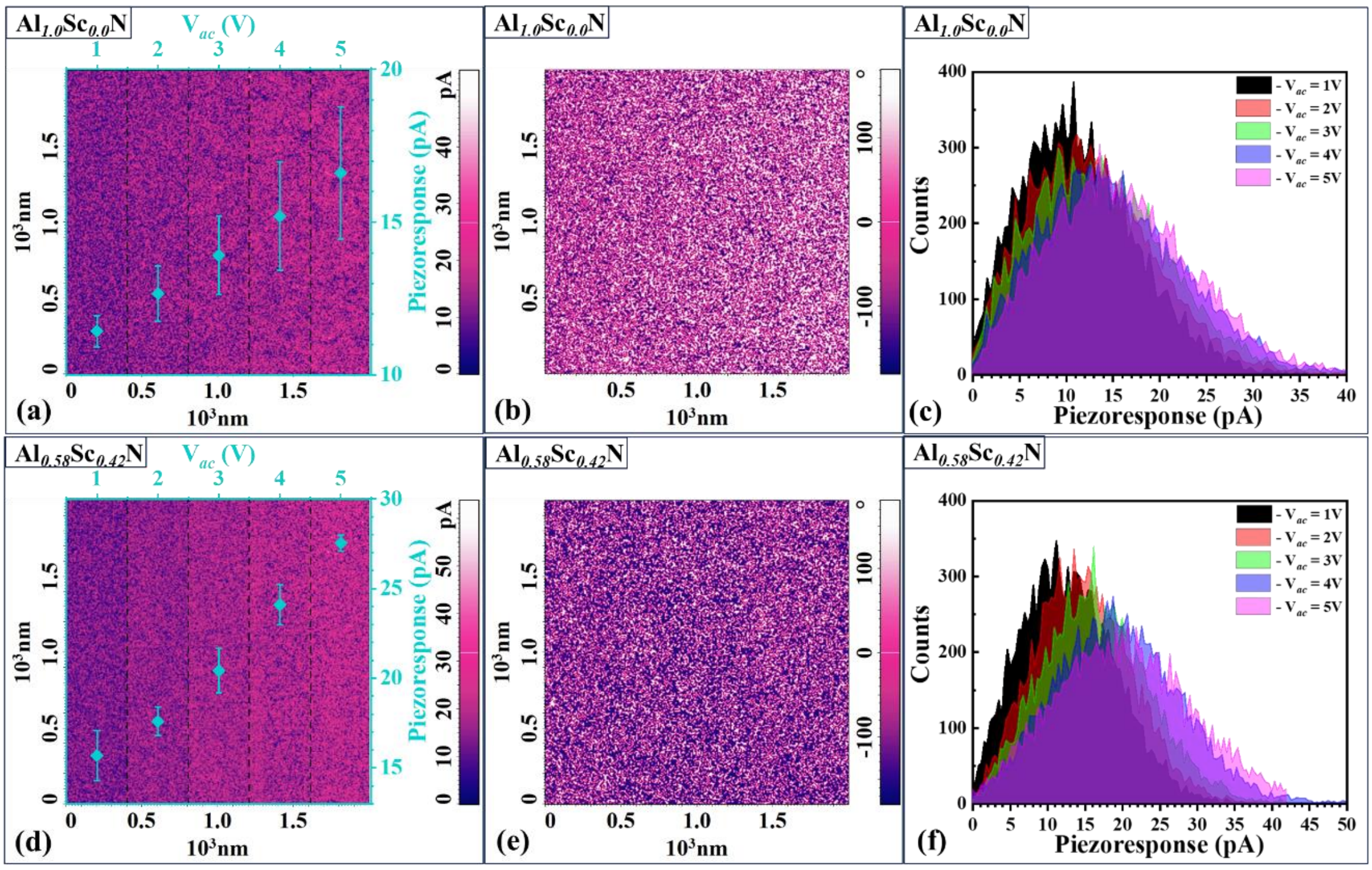

**Fig. 7:** Piezoresponse amplitude, phase image and histogram plots of: AlN thin film {(a) to (c)}, and $Al_{0.58}Sc_{0.42}N$ thin film {(e) to (f)}, obtained with a progressive sweeping of $V_{ac}$ using configuration-1 (*starting from 1 V at the left-most region of the scanned area to 5 V at the right-most region of the scanned area as indicated*)

Then, results for the second mode of data acquisition are shown in Fig. 7(a), (b) and (c), which respectively shows the piezoresponse amplitude (in pA), the corresponding phase images and the histogram plots, of the AlN thin film upon sweeping the $V_{ac}$ progressively from 1 V (left-most region of the scanned area) to 5 V (right-most region of the scanned area), respectively, as indicated in Fig. 7(a). Similarly, Fig. 7(e), (f), and (g), respectively, show the same set of results from the progressive $V_{ac}$-sweeping (left-to-right) of the $Al_{0.58}Sc_{0.42}N$ thin film. In this mode also, we could clearly observe a systematic increase in the piezoresponse amplitude with an increase in the value of the applied $V_{ac}$ from the amplitude images and their corresponding histograms. For the AlN thin film, the mean value of the histogram of the piezoresponse amplitude increases linearly from 11.7 pA at 1 V to 16 pA at 5 V. In the case of the $Al_{0.58}Sc_{0.42}N$ thin film, the mean value of histogram of the piezoresponse amplitude increases linearly from 15.4 pA at 1 V to 27.7 pA at 5 V. Then, several such $V_{ac}$-sweep measurements were performed (in both the modes for each configuration) on the specimen by changing the scanned area and scanning directions to arrive at a final $V_{ac}$ *vs*. piezoresponse amplitude plot like the ones embedded and shown in Fig. 7(a) and Fig. 7(d). Thereafter, the slope of the linear regression of the $V_{ac}$ *vs*. piezoresponse amplitude (in pm) plot will give us an estimate of the $d_{33\text{-}eff}$ of the specimen.

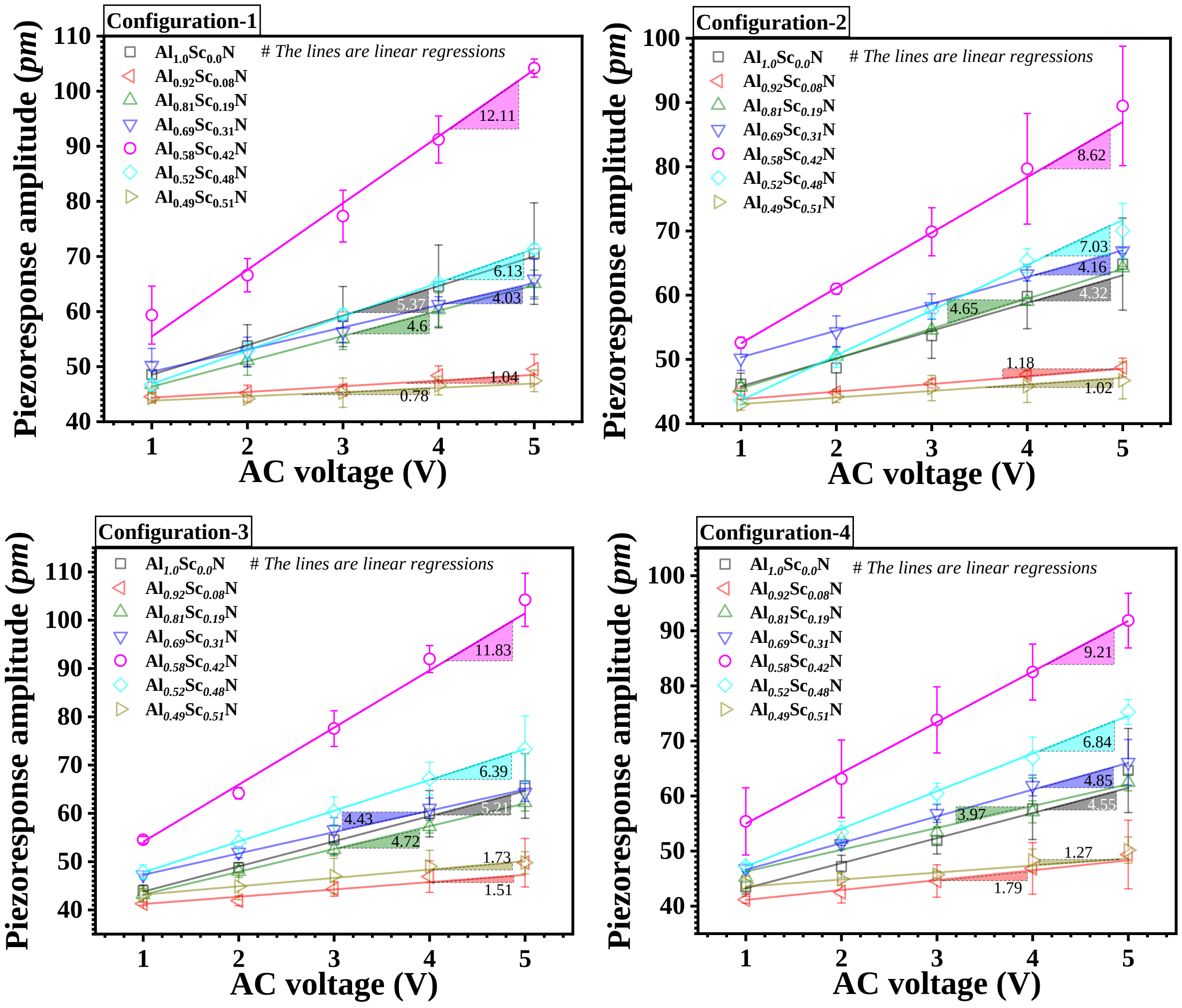

**Fig. 8:** Results of the linear regression on the $V_{ac}$ *vs*. piezoresponse amplitude plots for various $Al_{1-x}Sc_xN$ thin films in the four configurations described in Fig. 5.

The $V_{ac}$-sweep plots showing the piezoresponse amplitude of all the $Al_{1-x}Sc_xN$ thin film specimens using the above methodology in all the four different configurations described earlier are given in Fig. 8. The result of linear regressions performed on these $V_{ac}$-sweep of various $Al_{1-x}Sc_xN$ thin films are also illustrated in Fig. 8 for all the configurations along with the slopes ($d_{33\text{-}eff}$) obtained in each case. Prior to discussing the effect of Sc inclusion in the piezoresponse of the $Al_{1-x}Sc_xN$ thin films, we will discuss about the efficacy of the different configurations that we have used for measuring the piezoresponse. For this purpose, the values of $d_{33\text{-}eff}$, estimated independently in the four different configurations are plotted as a function of at. % of Sc in Fig. 9.

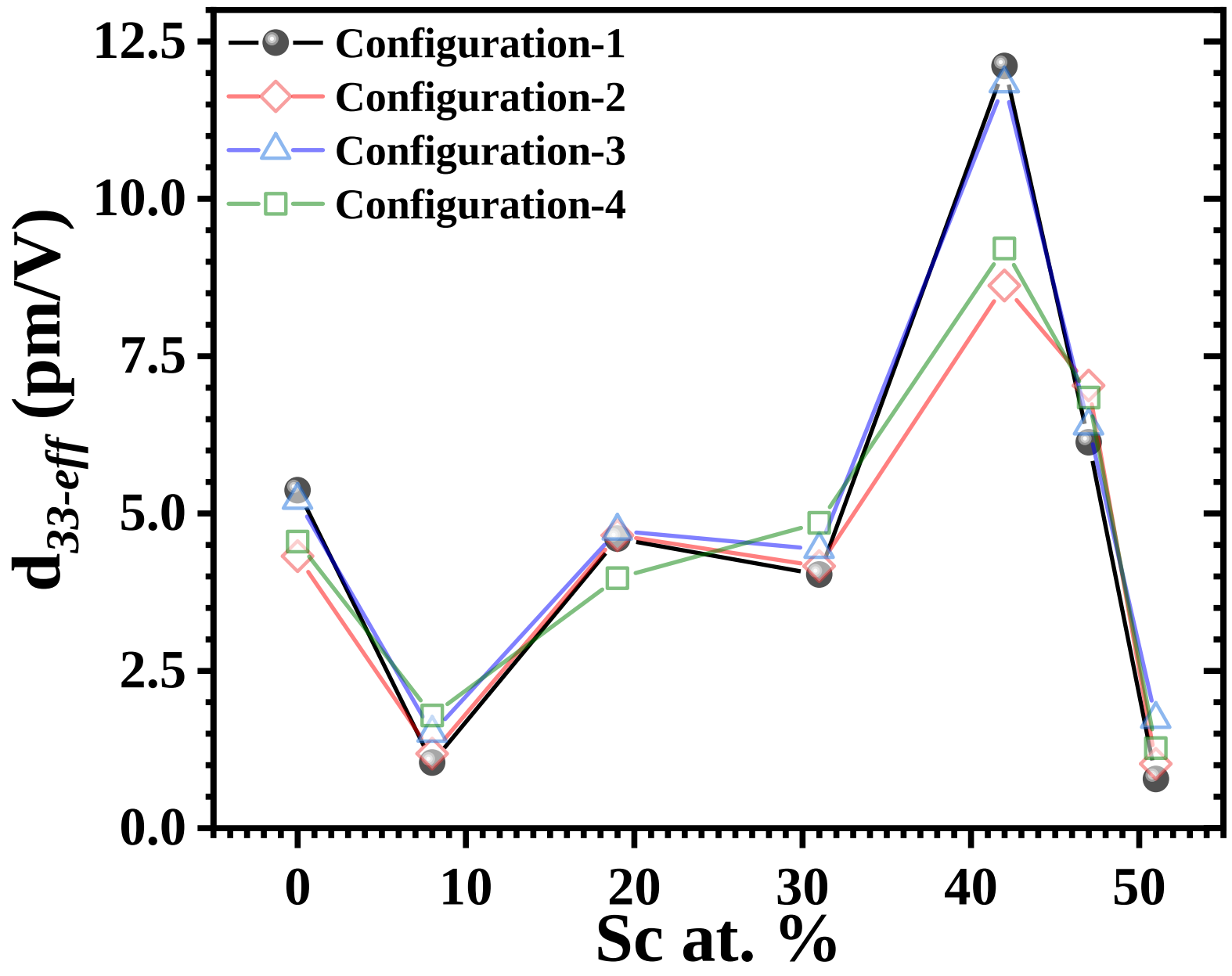

**Fig. 9:** Effective piezoelectric coefficient, $d_{33\text{-}eff}$ of sputtered deposited $Al_{1-x}Sc_xN$ thin films as a function of the at. % of Sc atoms

The plot shows that all the four configurations exhibit a similar variational trend in their values of $d_{33\text{-}eff}$ with respect to the at. % of Sc. However, the values of configuration-1 and 3, where no additional *dc*-offset is applied, are very closely matched and are precise to within 3 % (except ~ 6 % for $Al_{0.69}Sc_{0.31}N$, and higher for $Al_{0.92}Sc_{0.08}N$ and $Al_{0.49}Sc_{0.51}N$). We believe the high standard deviation in the last two samples mentioned above arises due to their inherent low values of $d_{33\text{-}eff}$ which is close to the background noise level of the lock-in amplifier used in our AFM. The close matching indicates the effectiveness of the method of positioning the laser spot of the OBB at $x$/L ~ 0.617 in minimizing the non-local electrostatic effects from the overhanging Si-chip. Then the set of $d_{33\text{-}eff}$ values obtained with the application of *dc*-offsets i.e., configuration-2 and 4 also closely matches within 2 to 11 %, except again for $Al_{0.92}Sc_{0.08}N$ and $Al_{0.49}Sc_{0.51}N$ for similar reason discussed above. Finally, when we compare the data from the two sets i.e., with (configuration-1 & 3) and without (configuration-2 & 4) *dc-offsets*, we observed a slight variation between the two data sets. The variation however, is random and does not exhibit a bias indicating that the local electrostatic effect have been minimized. Beyond this, there seems to be no plausible way either to reduce the effect further or even to tell if the effect has been perfectly minimized by the *dc*-offset, due to the uncertainties involves in the measurement. However, the two data sets exhibit a reasonable agreement with their values falling within a precision of ~ 10% for all the thin films except for slightly higher deviation of ~ 20 % for $Al_{0.58}Sc_{0.42}N$ thin film. In the case of $Al_{0.58}Sc_{0.42}N$ thin film, the application of *dc*-offset reduces the measured $d_{33\text{-}eff}$ value. Hence, in order to rationalize the overall data and improve precision, we have performed a statistical averaging of the values obtained using the four configurations (shown in Fig. 5) to estimate the final $d_{33\text{-}eff}$ values. The result thus obtained for our set of $Al_{1-x}Sc_xN$ thin films are tabulated in Table 1.

**Table 1:** Piezoelectric coefficient of the $Al_{1-x}Sc_xN$ thin films

| Sample | Crystal phase | Piezoelectric coefficient $d_{33\text{-}eff}$ (pm/V) |
|---|---|---|
| **$Al_{1.00}Sc_{0.00}N$ (AlN)** | *wurtzite* | 4.9 ± 0.5 |
| **$Al_{0.92}Sc_{0.08}N$** | *wurtzite* | 1.4 ± 0.3 |
| **$Al_{0.81}Sc_{0.19}N$** | *wurtzite* | 4.5 ± 0.4 |
| **$Al_{0.69}Sc_{0.31}N$** | *wurtzite* | 4.4 ± 0.4 |
| **$Al_{0.58}Sc_{0.42}N$** | *wurtzite* | 10.4 ± 1.8 |
| **$Al_{0.53}Sc_{0.47}N$** | *wurtzite phase mixed with cubic* | 6.6 ± 0.4 |
| **$Al_{0.49}Sc_{0.51}N$** | *cubic* | 1.2 ± 0.4 |

We now discussed the effect of Sc incorporation in the $d_{33\text{-}eff}$ of $Al_{1-x}Sc_xN$ thin films in the light of the structural changes induced by the addition of Sc in the *wz*-AlN lattice and compare these values with available literature data. Generally, piezoelectricity is not expected in a perfectly polycrystalline material as the polarizations or electric dipoles of all the randomly oriented domains exactly cancels each other resulting in a net-zero piezoresponse when an external $V_{ac}$ is applied (unless one pole the domains using a very high field like in the case of a ferroelectric material)[4]. However, when crystalline texture or preferential orientation occur in such a polycrystalline structure, the structural grains including the domain containing the electric dipoles restructure themselves to introduce anisotropy in an otherwise isotropic distribution of the electric dipoles. As a result, these dipoles will fail to exactly cancel out each other and we may get a net non-zero piezoelectric response in the direction of the applied $V_{ac}$ when such materials are probed. With this basis for the observation of piezoresponse in a polycrystalline material, we note that piezoelectricity in *wz*-AlN results from the spontaneous polarization of the *wz*-AlN structure along the *c*-axis due to the polar nature of the Al─N bonds [27]. We also recalled that our *wz*-AlN thin film exhibits a strong preferential orientation toward the (002) reflection i.e., the *c*-axis (Fig. 1), the natural direction for polarization in wurtzite structure. This explain the high value of $d_{33\text{-}eff}$ ~ 4.9 pm/V obtained for our *wz*-AlN thin film which is comparable with the values of other *c*-axis oriented AlN thin film reported in literatures as shown in Fig. 10.

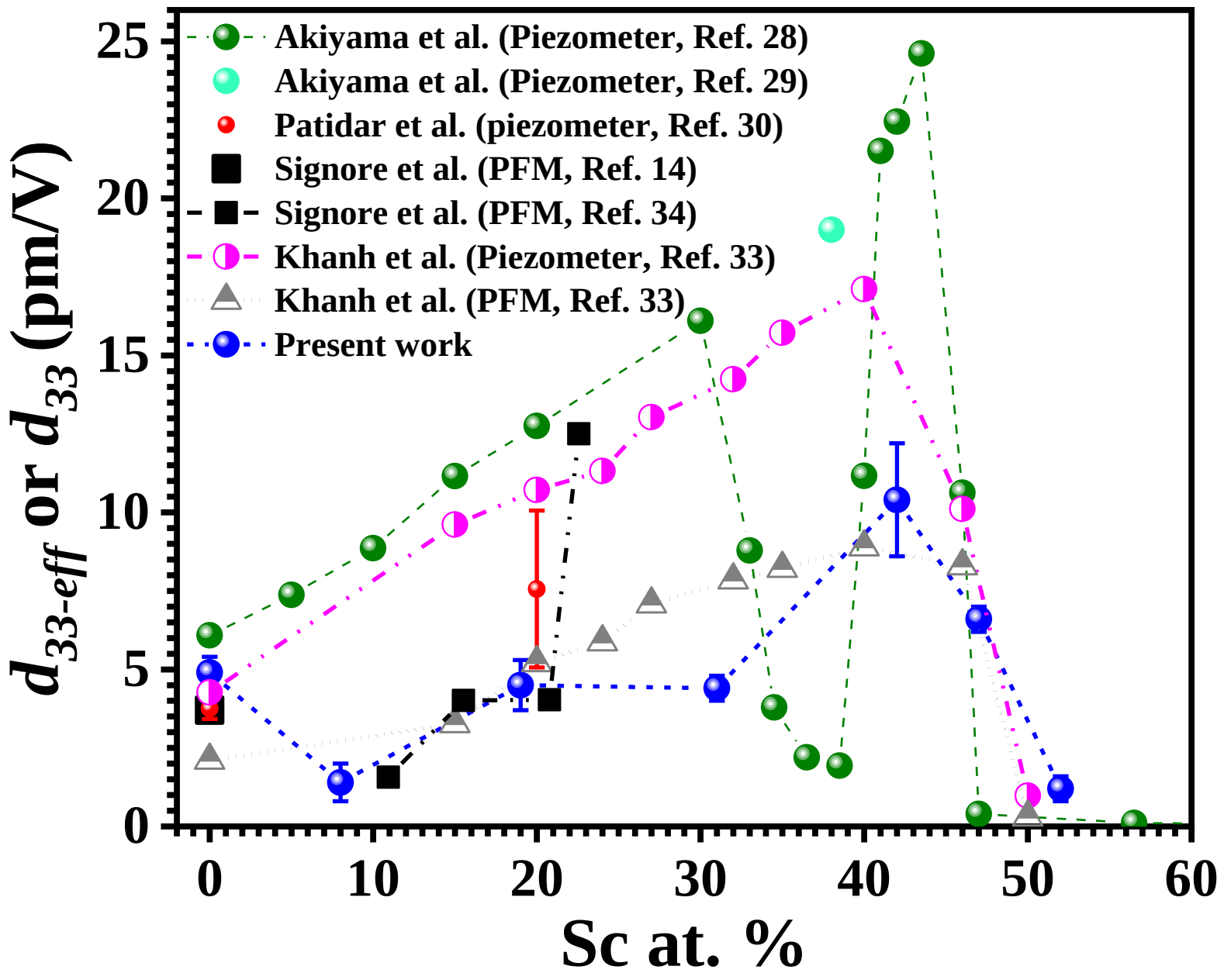

**Fig. 10:** Comparison of the effective piezoelectric coefficient, $d_{33\text{-}eff}$ of our $Al_{1-x}Sc_xN$ thin films with values reported in literature[14,28–32]. (Note: compositional data given in Ref. 34 has been normalized to present the at. % of Sc in the $Al_{1-x}Sc_xN$ form for easy comparison).

Thereafter, the process of alloying Sc into the *wz*-AlN to realize *wz*-$Al_{1-x}Sc_xN$ alloy thin films result in the destruction of the *c*-axis orientation that exist in the *wz*-AlN, particularly at low at. % of Sc (Fig. 1). The cationic substitution of Al with Sc (cation with a larger ionic radius and electropositivity) increase the ionicity of the Al─N bond and causes strain-induced distortion of the *wz*-AlN lattice, similar to the case of Cr-alloying in *wz*-AlN [33,34]. While the distortion is such that the *c*-axis undergo compression along with an expansion of the basal plane, the effect of increased ionicity is felt the most along the *c*-axis [27]. The compression of *c*-axis and expansion of *a*-axis can be inferred from the shifts of the (001) and (002) peak positions, respectively, toward lower and higher 2θ values in the XRD pattern (Fig. 1). Although these shifts indicative of the distortion is present in the $Al_{0.92}Sc_{0.08}N$ thin film, the lack of any preferred orientation along the *c*-axis due an almost perfect polycrystallinity is responsible for its low $d_{33\text{-}eff}$ ~ 1.4 pm/V. Then, as $x$ increases, the shifts in the XRD peak positions and hence, the distortional strain in the *wz*-$Al_{1-x}Sc_xN$ lattice increases (Fig. 1). Here, we also highlight that the expansion of the basal plane (*a*-axis) is mostly linear and systematic, whereas the compression of the *c*-axis is not linear or systematic. As a result, the *wz*-$Al_{1-x}Sc_xN$ thin films are observed to exhibit a varying degree of preferential orientation along the *c*-axis (Fig. 1) with increasing at. % of Sc. Accordingly, the $Al_{0.81}Sc_{0.19}N$ and $Al_{0.69}Sc_{0.31}N$ thin films exhibit $d_{33\text{-}eff}$ values of ~ 4.5 pm/V. We further observed that the $Al_{0.58}Sc_{0.42}N$ thin film accommodate the maximum strain while still maintaining the phase purity of the *wurtzite* lattice. This led to the maximum observed $d_{33\text{-}eff}$ value of ~ 10.4 pm/V for the $Al_{0.58}Sc_{0.42}N$ among the set of $Al_{1-x}Sc_{x}N$ thin films. Any further increase in $x$ beyond 0.42 led to phase mixing of the wurtzite phase with cubic $Sc_3AlN$ phase. The mixing of phase led to a decrease in the value of $d_{33\text{-}eff}$ until at $x \approx 0.51$, the alloy crystallizes only in the cubic-$Sc_3AlN$ phase. The inversion symmetry introduced by this cubic ternary alloy phase (space group Pm-3m) destroy any piezoelectricity in the sample and the $d_{33\text{-}eff}$ of the $Al_{0.49}Sc_{0.51}N$ reduce to ~ 1.2 pm/V at the baseline level. The composition that maximizes the enhancement in the $d_{33\text{-}eff}$ of the $Al_{1-x}Sc_{x}N$ thin film agrees well with earlier reports as can be seen in Fig. 10 [14,28–32]. However, the $d_{33\text{-}eff}$ of our polycrystalline-$Al_{0.58}Sc_{0.42}N$ thin film is observed to be less than the values reported for similar composition, measured mostly using a piezometer set-up. The reduction may be attributed to the lack of strong orientation or texturing along the *c*-axis in our $Al_{0.58}Sc_{0.42}N$ thin film due to the degree of polycrystallinity achieved with the deposition condition used in the present study. It is also compounded by the fact that no correction for substrate clamping effect has been considered in our study. Nevertheless, we feel that the $d_{33\text{-}eff}$ can be increased by realizing either a pseudo-single crystalline or a perfect *c*-axis oriented $Al_{0.58}Sc_{0.42}N$ thin film by optimizing the deposition and post-deposition process parameters. Such optimized $Al_{0.58}Sc_{0.42}N$ thin films can find good application in actuators and SAW devices that needs high temperature stability and will be a topic of future studies.

## Conclusion

We have presented a methodology that aim to improve the quantitative determination of the effective piezoelectric coefficients ($d_{33\text{-}eff}$) of a polycrystalline piezoelectric thin film. A combination of experimental measures is used to either eliminate or minimize the parasitic electrostatic contribution that usually plague the quantitative characterization of $d_{33\text{-}eff}$. These measures include optimizing the sensitivity of the optical lever to *dc* or electrostatic voltages, applying additional *dc*-offset during PFM scans, and using a scanning position that minimizes tip overhang over the sample and the corresponding formation of image charges. The above methodology was used the estimate the $d_{33\text{-}eff}$ of a set of Sc alloyed AlN ($Al_{1-x}Sc_{x}N$) thin films. The pristine AlN thin film crystallizes in the wurtzite (*wz*) phase and exhibit a preferred orientation along the (002) reflection or the *c*-axis. The $d_{33\text{-}eff}$ of the *wz*-AlN thin film was estimated to be ~ 4.9 pm/V, consistent with reported values of other *c*-axis oriented AlN thin film. While the addition of Sc into *wz*-AlN reduce the texturing along the *c*-axis in the present growth conditions, the highest at. % of Sc that still retain the phase purity of the wurtzite phase and also shows the maximum enhancement in the $d_{33\text{-}eff}$ value for $x \approx 0.42$. Above this, phase-mixing of the *wz*-phase with the *cubic*-$Sc_3AlN$ phase sets in and $d_{33\text{-}eff}$ reduces till it become ~ 1.2 pm/V at $x \approx 0.51$.

**Acknowledgements**

The authors would like to acknowledge UGC-DAE CSR for extending the XRD and DC sputtering facilities. The authors thank Mrs. D.T. Sunitha Rajkumari for her help during the GIXRD measurements. The authors thank Dr. Sandip Dhara, Director, MSG/IGCAR for careful reading and suggesting valuable corrections in the manuscript. The authors also thank Director, Indira Gandhi Centre for Atomic Research, Kalpakkam for encouragement and support during this work.